\documentclass[aps,preprint,prb]{revtex4}
\usepackage{graphicx}
\usepackage{epsfig}
\usepackage{amsmath}
\usepackage{graphics}
\begin{document}

\author{M. F. Gelin}
\author{D. S. Kosov}

\affiliation{Department of Chemistry and Biochemistry,
 University of Maryland, 
 College Park, 
 MD 20742 }

\title{What can be learned about molecular reorientation from single molecule polarization microscopy?}

\begin{abstract}
We have developed a general approach for the calculation of the single
molecule polarization correlation function $C(t)$, which delivers
a correlation of the emission dichroisms at time $0$ and $t$. The
approach is model independent and valid for general asymmetric top
molecules. The key dynamic quantities of our analysis are the even-rank
orientational correlation functions, the weighted sum of which yields
$C(t)$. We have demonstrated that the use of non-orthogonal schemes
for the detection of the single molecule polarization responses makes
it possible to manipulate the weighting coefficients in the expansion
of $C(t)$. Thus valuable information about the orientational correlation
functions of the rank higher than second can be extracted from $C(t)$. 
\end{abstract}
\maketitle

\section{Introduction }

The formulation of the polarization-sensitive spectroscopy in terms
of the orientational correlation functions (OCFs) has opened up a
possibility of the unified description of various signals and clarified
information content of different spectroscopic techniques. \cite{gor68}
As has been demonstrated, all large variety of different polarization
signals can be described by the OCFs of the first and second rank.
\cite{gor68,McCl77,BurTe,zew01} More specifically, one can ''measure''
either OCFs in the time domain, or their time derivatives, or their
Fourier spectra, or their integral relaxation times. It is this unification
which has made it possible to compare the results of different measurements
and to learn about the mechanisms of molecular reorientation, both
in the gas phase and in the condensed phase. The contemporary nonlinear
(third order) ultrafast polarization spectroscopy is interpreted in
terms of the three-time correlation functions (CFs) of the dipole
moments or polarizability tensors involved. \cite{muk1} However,
in practices, their influence on a signal is normally accounted for
by a static averaging, or the three-time polarization CFs reduce to
the standard OCFs of the second rank, due to either the strong optical
dephasing or the shortness of the laser pulse on the rotational dynamic timescale. 

All the written above pertains to spectroscopy of ensembles, in
which the measured response is averaged over many single-particle
contributions. The situation with the single molecule (SM) spectroscopy
in general, \cite{orr99,orr04,sti04,osa06,buc06} and the SM polarization spectroscopy
in particular, \cite{wei99,orr04} is very different. The SM
signal delivers a response of an individual system, whose time dependence
reports about fluctuations caused by the system "nanoenvironment''.

Ideally, one would prefer to measure the three-dimensional orientation of the emission dipole moment(s) in real time. Indeed, there exist several schemes which allow us to do that\cite{orr04,moe98,nov00,nov04,end03,kot03,hof06}. These  techniques require, typically,  many photons to get a good signal-to-noise ratio. This obstacle restricts the length and the time resolution of the recorded signal. 
Fourkas has suggested that  the time evolution of the emission dipole can be determined "on the fly", by detecting  SM emission along three different polarization directions.\cite{fou01} Hohlbein and  H\"{u}bner have recently implemented this method.\cite{hub05}

So far, many SM experiments have been designed to detect the in-plane projection of the SM emission
along two different (mutually perpendicular) polarization directions.  Thereby  long enough transients can be measured,
which permit of the reliable calculation of SM CFs.
The SM dichroism CF is the key dynamic quantity which is delivered by the polarization SM fluorescence microscopy.\cite{hof06,bou02,hoc03,rei06,bas04,bou05} This CF is much more complicated than its counterparts which describe polarization responses in ensemble measurements.

The SM dichroism CF is the core object of the present study, 
which has been inspired by the recent papers.\cite{bas04,bou05}  
Our aim is threefold. First, we generalize the approach, which has been
developed in \cite{bas04,bou05} for spherical top molecules within
the small-angle rotational diffusion model, to asymmetric top molecules
and beyond any particular model of molecular reorientation. The key
dynamic quantities here are the even-rank OCFs, which can either be
evaluated within any model of the molecular reorientation available
in the literature, \cite{McCl77,BurTe,kos06} or simulated on a computer.
\cite{AlTi} Second, we discuss a possibility of gaining additional
knowledge about molecular reorientation by utilizing non-orthogonal
schemes for the detection of SM polarization responses. Third, we
demonstrate that a valuable information about the OCFs of the rank
higher than second can be extracted from the SM CFs. Until recently,
such an information was available only through the computer simulations
and model calculations.

\section{Single molecule dichroism}

The key quantity in the SM polarization microscopy is the dichroism
\begin{equation}
X\equiv\frac{I_{1}-I_{2}}{I_{1}+I_{2}}.
\label{X}
\end{equation}
 Here $I_{i}$ are the intensities of the light emission which are
detected at two different (usually, mutually perpendicular) polarizations
$\mathbf{e}_{i}$, $i=1,\,2$. Within the oscillator model, the excitation
and emission processes are independent, so that the emission intensity
is proportional to the product of two probabilities, \cite{gor68,feo}
\begin{equation}
I_{i}(t)\sim\sigma_{a}(0)\sigma_{i}(t),\label{I}\end{equation}
and the SM  dichroism  (\ref{X}) is independent of the absorption cross-section $\sigma_a$.
The absorption probability is given by the square of the scalar
product of the polarization of the absorbed light $\mathbf{e}_{a}$
and the absorption dipole moment $\mu_{a}$: \cite{foot2}
\begin{equation}
\sigma_{a}\sim(\mathbf{e}_{a}\mathbf{\mu}_{a})^{2}.
\label{Sa}
\end{equation}
 The emission cross-section is given by a similar expression, \cite{agr}
 \begin{equation}
\sigma_{i}\sim(\mathbf{e}_{i}\mathbf{\varepsilon})^{2},
\label{Si}
\end{equation}
in which the polarization of the emitted light, $\mathbf{\varepsilon}$,
is explicitly given by the equation\begin{equation}
\mathbf{\varepsilon}=\mathbf{k} \times \mathbf{\mu} \times \mathbf{k} =   \mathbf{\mu}-\mathbf{\mathbf{k}}(\mathbf{\mu\mathbf{k}}),\label{eps}\end{equation}
$\mathbf{k}$ being the unit vector along the propagation of the light beam,
and $\mathbf{\mu}$ being the emission dipole moment. \cite{foot3}

When the
fluorescence is collected from an ensemble of molecules, molecular  contributions add
up incoherently. Thus, in order to get the total emission intensity,
one has to average Eq.(\ref{eps}) over all possible orientations
of the wave vector $\mathbf{k}$. This yields $\left\langle \mathbf{\varepsilon}\right\rangle _{\mathbf{k}}\sim\mathbf{\mu},$
so that the averaged emission cross-section becomes similar to its
absorption counterpart:\begin{equation}
\sigma_{i}\sim(\mathbf{e}_{i}\mathbf{\mu})^{2}.\label{SiEns}\end{equation}
Eq.(\ref{I}), in conjunction with Eqs.(\ref{Sa}) and (\ref{SiEns}),
embody the standard starting point for the calculation of the intensity
of the polarized emission in ensemble measurements, $\left\langle I_{i}(t)\right\rangle  \sim \left\langle \sigma_{a}(0)\sigma_{i}(t) \right\rangle$, see Refs. \cite{gor68,feo}

If we collect emission from a single molecule, the above incoherent
averaging procedure is not legitimate any longer. Since the wave vector
$\mathbf{k}$ is specific to a photon which has been emitted by the
molecule but the direction of $\mathbf{k}$ is unknown, we have to
average the emission probability (\ref{Si}) over all those $\mathbf{k}$
which can be collected by the objective. Let $\mathbf{s}$ be the
unit vector along the axis of the objective ($\mathbf{s}\bot\mathbf{e}_{i})$
and $\vartheta_{\mathbf{k}},\,\phi_{\mathbf{k}}$ be the spherical
angles of the unit vector $\mathbf{k}$, so that $\mathbf{ks}=\cos(\vartheta_{\mathbf{k}})$.
We also introduce the light-collection cone angle \begin{equation}
\chi=\arcsin(NA/n),\label{NA}\end{equation}
 $NA$ being the numerical aperture of the objective and $n$ being
the refraction index of the medium in which the sample is embedded 
(Fig. 1 clarifies the meaning of the introduced quantities).
If we further assume that the emitting molecule is located at the
focal point of the ideal and polarization-preserving objective which
collects the emission, then the $\mathbf{k}$-averaging of the emission
probability (\ref{Si}) yields\begin{equation}
\overline{\sigma}_{i}\equiv\int_{0}^{2\pi}d\phi\int_{-\chi}^{\chi}\sin\vartheta d\vartheta\sigma_{i}\sim p+q(\mathbf{s\mu})^{2}+2(\mathbf{e}_{i}\mathbf{\mu})^{2}.\label{SiSi}\end{equation}
The numerical parameters $p$ and $q$ are uniquely determined by
the collection angle $\chi$: \cite{wei99,bas04,bou05,axe79,fou01}\[
p=(A+B)/C-1,\,\, q=1-B/C,\]
where the quantities $A,\, B,\, C$ are defined in the standard way:
\begin{equation}
A=8-12\cos(\chi)+4\cos^{3}(\chi),\label{A}\end{equation}
\begin{equation}
B=6\cos(\chi)-6\cos^{3}(\chi),\label{B}\end{equation}
\begin{equation}
C=7-3\cos(\chi)-3\cos^{2}(\chi)-\cos^{3}(\chi).\label{C}\end{equation}
Eq.(\ref{SiSi}) generalizes slightly its counterparts from,\cite{wei99,bas04,bou05,axe79,fou01}
allowing for an arbitrary direction of the polarizer $\mathbf{e}_{i}\bot\mathbf{s}$
in the laboratory frame. 

Fig.2 shows the behavior of the coefficients of $p(\chi)$ and $q(\chi)$. Both of them increase monotonically with $\chi$: $p(0)$$=q(0)$$=0$,
while $p(\pi/2)=1/7$ and $q(\pi/2)=1$. For every $\chi$, $q(\chi)>p(\chi)$.
Eq.(\ref{SiSi}) reduces to Eq.(\ref{SiEns})
in the limit of small collection angle ($\chi\rightarrow0$). In this
case, the beams with $\mathbf{k}\Vert\mathbf{s}$ are collected only,
the $\mathbf{k}$-dependent portion of Eq.(\ref{eps}) does not contribute
into Eq.(\ref{Si}), so that Eq.(\ref{SiSi}) reduces to (\ref{SiEns})
since both $p\rightarrow0$ and $q\rightarrow0$. 

After the insertion of Eq.(\ref{SiSi}) into Eq.(\ref{X}), one
gets the general expression for the SM dichroism:
\begin{equation}
X=\frac{(\mathbf{e}_{1}\mathbf{\mu})^{2}-(\mathbf{e}_{2}\mathbf{\mu})^{2}}{p+q(\mathbf{s\mu})^{2}+(\mathbf{e}_{1}\mathbf{\mu})^{2}+(\mathbf{e}_{2}\mathbf{\mu})^{2}}.\label{Xgen}\end{equation}
If $NA=0$, then $p=q=0$ and the dichroism attains the familiar form
\begin{equation}
X=\frac{(\mathbf{e}_{1}\mathbf{\mu})^{2}-(\mathbf{e}_{2}\mathbf{\mu})^{2}}{(\mathbf{e}_{1}\mathbf{\mu})^{2}+(\mathbf{e}_{2}\mathbf{\mu})^{2}}.\label{Xsim}\end{equation}
If we employ the orthogonal signal detection scheme ($\mathbf{e}_{1}\bot\mathbf{e}_{2})$,
then Eq.(\ref{Xgen}) simplifies to \begin{equation}
X=\frac{(\mathbf{e}_{1}\mathbf{\mu})^{2}-(\mathbf{e}_{2}\mathbf{\mu})^{2}}{1+p-(1-q)(\mathbf{s\mu})^{2}}.\label{Xper}\end{equation}
 If, additionally, $NA$ is high ($\chi\rightarrow\pi/2$, $p(\pi/2)=1/7$
and $q(\pi/2)=1$), then the denominator in Eq.(\ref{Xper}) becomes
constant and \begin{equation}
X=\frac{7}{8}\{(\mathbf{e}_{1}\mathbf{\mu})^{2}-(\mathbf{e}_{2}\mathbf{\mu})^{2}\}.\label{Xsimp}\end{equation}

\section{Orientational correlation functions}

Let $\mathbf{\Omega}$ denote collectively the set of three Euler
angles $\alpha,\beta,\gamma$ which specify orientation of the molecular
frame with respect to the laboratory one. Let us further introduce
the conditional probability density function, \begin{equation}
\rho(\mathbf{\Omega}_{0}|\mathbf{\Omega},t),\label{Cond}\end{equation}
which is the probability density that the molecule has orientation
$\mathbf{\Omega}$ at time $t$, provided it had orientation $\mathbf{\Omega}_{0}$
at $t=0$. By definition, the quantity (\ref{Cond}) obeys the initial
condition $\rho(\mathbf{\Omega}_{0}|\mathbf{\Omega},0)=\delta(\mathbf{\Omega}-\mathbf{\Omega}_{0})$.
We also assume that the molecule can be subjected to an external (anisotropic)
potential $U(\mathbf{\Omega})$, so that the corresponding equilibrium
Boltzmann distribution reads

\begin{equation}
\rho_{U}(\mathbf{\Omega})=Z_{U}\exp\{-U(\mathbf{\Omega})/(k_{B}T)\},\label{U}\end{equation}
 $k_{B}$ being the Boltzmann constant, $T$ being the temperature,
and $Z_{U}$ being the partition function. Evidently, $\rho(\mathbf{\Omega}_{0}|\mathbf{\Omega},t\rightarrow\infty)=\rho_{U}(\mathbf{\Omega})$.

We are in a position now to define the OCF \cite{cuk74,zan91,zan93}
\begin{equation}
G_{ll';mm'}^{jj'}(t)\equiv\left\langle D_{lm}^{j}(\mathbf{\Omega}(0))D_{l'm'}^{j'*}(\mathbf{\Omega}(t))\right\rangle,\label{OCF}\end{equation}
$D_{lm}^{j}(\mathbf{\Omega})$ being the Wigner D-functions. \cite{var89}
If the conditional probability density function (\ref{Cond}) is known,
then the OCF (\ref{OCF}) can be evaluated as follows: 

\begin{equation}
G_{ll';mm'}^{jj'}(t)\equiv\int d\mathbf{\Omega}d\mathbf{\Omega}_{0}D_{lm}^{j}(\mathbf{\Omega})D_{l'm'}^{j'*}(\mathbf{\Omega}_{0})\rho_{U}(\mathbf{\Omega}_{0})\rho(\mathbf{\Omega}_{0}|\mathbf{\Omega},t).\label{Green}\end{equation}
The OCFs (\ref{OCF}) have been explicitly computed, for example,
within the rotational diffusion model. \cite{zan91,zan93} 

If the OCFs (\ref{OCF}) are known, one can evaluate any polarization response of interest. Indeed, the CF of any (for simplicity, real) orientation-dependent quantity
$B(\mathbf{\Omega})$ at the time moments $0$ and $t$ can immediately
be expressed through OCFs (\ref{OCF}): \begin{equation}
\left\langle B(\mathbf{\Omega}(0))B(\mathbf{\Omega}(t))\right\rangle =\sum_{j,j'=0}^{\infty}\,\sum_{l,m=-j}^{j}\,\sum_{l',m'=-j'}^{j'}B_{lm}^{j}G_{ll';mm'}^{jj'}(t)B_{l'm'}^{j'*}.\label{B(t)}\end{equation}
 Here \begin{equation}
B_{lm}^{j}\equiv\frac{2j+1}{8\pi^{2}}\int d\mathbf{\Omega}D_{lm}^{j}(\mathbf{\Omega})B(\mathbf{\Omega})\label{BD}\end{equation}
are the expansion coefficients of the quantity $B(\mathbf{\Omega})$
over the D-functions. 

There exists an important particular case, in which CF (\ref{B(t)})
simplifies greatly. Namely, let us assume that there are no external
fields ($U(\mathbf{\Omega})=0$), so that rotational phase space is
isotropic. Then one can show (see, e.g., Ref. \cite{cuk74} for more
details), that the fundamental OCF (\ref{OCF}) becomes \begin{equation}
G_{ll';mm'}^{jj'}(t)=\frac{1}{2j+1}\delta_{jj'}\delta_{ll'}G_{m'm}^{j}(t),\label{Green}\end{equation}
$\delta_{jj'}$ being the Kronecker symbol. Therefore, the CF (\ref{B(t)})
is now evaluated as \begin{equation}
\left\langle B(\mathbf{\Omega}(0))B(\mathbf{\Omega}(t))\right\rangle =\sum_{j=0}^{\infty}\,\sum_{m,m'=-j}^{j}A_{mm'}^{j}G_{m'm}^{j}(t),\label{B1(t)}\end{equation}
\begin{equation}
A_{mm'}^{j}=\frac{1}{2j+1}\sum_{k=-j}^{j}B_{km}^{j}B_{km'}^{j*}\label{AG}\end{equation}
(compare with \cite{val73,cuk72,pec76}). The quantities 

\begin{equation}
G_{mm'}^{j}(t)\equiv\sum_{k=-j}^{j}\left\langle D_{mk}^{j}(\mathbf{\Omega}(t))D_{m'k}^{j*}(\mathbf{\Omega}(0))\right\rangle \equiv\left\langle D_{mm'}^{j}(\mathbf{\Omega_{\Delta}}(t))\right\rangle .\label{OCF1}\end{equation}
 are the standard OCFs of the rank $j$ ($\mathbf{\Omega_{\Delta}}(t)$ being
the angle of relative reorientation). They can either be simulated
on a computer \cite{AlTi} or evaluated within the models of molecular
reorientation available in the literature. \cite{McCl77,BurTe,kos06}

Eq.(\ref{B1(t)}) can further be simplified in the following important
particular case. Let us assume that the quantity of interest, $B(\mathbf{\Omega})$,
is specified by the orientation of a unit vector (for example, a dipole moment) $\mathbf{d}$, which
is "rigidly attached'' to the molecule, i.e., $B(\mathbf{\Omega})\rightarrow B(\mathbf{d})$.
The orientation of the unit vector $\mathbf{d}$ is uniquely determined
by the the spherical angles $\alpha,\beta$ and is specified,
therefore, by the reduced D-function $D_{l0}^{j}(\alpha,\beta,0)$
which, apart from the numerical factor, coincides with the corresponding
spherical harmonics, \cite{var89} 
\begin{equation}
Y_{jl}(\alpha,\beta)=\sqrt{\frac{2j+1}{4\pi}} D_{l0}^{j}(\alpha,\beta,0)\label{Dred}
\end{equation}
On writing Eq.(\ref{Dred}), we
have tacitly assumed that the unit vector $\mathbf{d}$ is pointed
along the $z$-axis of the molecular reference frame. In this case,
the terms with $m=m'=0$ survive only in Eq.(\ref{B1(t)}), so that
\begin{equation}
\left\langle B(\mathbf{d}(0))B(\mathbf{d}(t))\right\rangle =\sum_{j=0}^{\infty}\, A_{00}^{j}G_{00}^{j}(t).\label{B2(t)}\end{equation}
Thus, from the formal point of view, the CF (\ref{B2(t)}) is determined
by the linear combination of the OCFs $G_{00}^{j}(t)$ with different
ranks $j$. Of course, the relative significance of the contributions
from different OCFs is determined by the weighting coefficients $A_{00}^{j}$,
which depend on a particular form of the quantity under study, $B(\mathbf{d})$.

\section{Single molecule signal through orientational correlation functions}

The CF, which is normally extracted from the SM polarization signal,
is defined via the expression \cite{bou02,hoc03,bas04,bou05}
\begin{equation}
C(t)=\left\langle X(\mathbf{\Omega}(0))X(\mathbf{\Omega}(t))\right\rangle, \label{C(t)}\end{equation}
in which $X(\mathbf{\Omega})$ is
the SM dichroism (\ref{Xgen}) and its time-dependence is induced by
molecular rotation. \cite{foot1} To apply a general formalism outlined in the previous section 
for the evaluation of Eq.(\ref{C(t)}), one has to express all the
scalar products in Eq.(\ref{Xgen}) in terms of the Wigner D-functions. This is easily achieved by the formula
\begin{equation}
(\mathbf{s\mu})^{2}=\frac{1}{3}\left(1+2\sum_{l,m=-2}^{2}D_{0l}^{2}(0,-\beta_{s},-\alpha_{s})D_{lm}^{2}(\mathbf{\Omega})D_{m0}^{2}(\alpha_{\mu},\beta_{\mu},0)\right)\label{ms}\end{equation}
and similar expressions for the scalar products $ (\mathbf{e}_{i}\mathbf{\mu})^{2}$. Emphasize that the spherical angles $ \alpha_{s},\beta_{s}$ and $\alpha_{\mu},\beta_{\mu}$ in Eq.(\ref{ms}) are time-independent and specify orientations of the vectors $\mathbf{s}$ and $\mathbf{\mu}$, correspondingly,     
in the laboratory and molecular reference frames. The proper description of molecular reorientation is accounted for by the (time-dependent) Euler angles $\mathbf{\Omega}$.

To simplify the subsequent calculations, we can proceed as has been described in the previous section and  choose the molecular reference frame in such a way that
the emission dipole moment $\mathbf{\mu}$ is directed along the $z$-axis
of this frame. This is tantamount to putting $\alpha_{\mu}=\beta_{\mu}=0$ in Eq.(\ref{ms}). 
Then $D_{m0}^{j}(0,0,0)=\delta_{m0}$ (the
latter being the Kronecker delta), and Eq.(\ref{ms})
becomes independent of the angle $\gamma$ of rotation about the molecular
$z$-axis:
\begin{equation}
(\mathbf{s\mu})^{2}=\frac{1}{3}\left(1+2\sum_{l=-2}^{2}D_{0l}^{2}(0,-\beta_{s},-\alpha_{s})D_{l0}^{2}(\alpha,\beta,0)\right).\label{ms1}\end{equation}
Upon the insertion of this formula (and similar expressions for $ (\mathbf{e}_{i}\mathbf{\mu})^{2}$) into Eq.  
(\ref{BD}) one realizes that dichroism $X$ (Eq.(\ref{Xgen})) becomes a function of two Euler angles  $\alpha$ and $\beta$, $X(\alpha,\beta)$. Thus, Eqs. (\ref{BD})-(\ref{OCF1}) are applicable to this case. By  using the interconnection (\ref{Dred}) between the spherical harmonics and D-functions, one finally gets 
\begin{equation}
C(t)=\sum_{j=0}^{\infty}A_{00}^{j}G_{00}^{j}(t),\,\,\, A_{00}^{j}=\sum_{k=-j}^{j}|X_{k0}^{j}|^{2},\label{C(t)sim}\end{equation}
\begin{equation}
X_{k0}^{j}\equiv \sqrt{\frac{1}{4\pi}}\int_{0}^{\pi}\sin\beta d\beta \int_{0}^{2\pi} d\alpha \, Y_{jk}(\alpha,\beta,0)X(\alpha,\beta).\label{BD1}\end{equation}
These are exactly the formulas which have been derived in Refs.\cite{bas04,bou05}
Emphasize that the domain of validity of Eqs. (\ref{C(t)sim}), (\ref{BD1}) is
not confined to linear rotors and spherical tops. They are valid
for a general asymmetric top molecule and will be employed in the
remainder of the present paper. 

A word of caution is however in order. 
First, one should keep in mind that Eqs. (\ref{C(t)sim}), (\ref{BD1}) are applicable for molecular rotation in  an isotropic space. If there exist external potentials, one should use a more general formula (\ref{B(t)}). 
Second, Eqs. (\ref{C(t)sim}), (\ref{BD1}) are valid
if and only if the molecular $z$-axis coincides with the direction
of the emission dipole moment $\mathbf{\mu}$. This direction, however,
may not coincide with the molecular symmetry axis. This means that
such a choice of the molecular frame may not accommodate the molecular
symmetry properly. In order to do that and to evaluate $G_{00}^{j}(t)$
efficiently, we can switch from the initial frame to another "convenient"
molecular frame via the corresponding Wigner matrix $D_{mk}^{j}(-\mathbf{\Delta})$,
evaluate the OCF $\widetilde{G}_{kn}^{j}(t)$ in the reference frame
which fully accounts for molecular symmetry and return back to the
original molecular frame:\begin{equation}
G_{00}^{j}(t)=\sum_{k,n=-j}^{j}D_{0k}^{j}(-\mathbf{\Delta})\widetilde{G}_{kn}^{j}(t)D_{n0}^{j}(\mathbf{\Delta}).\label{MM}\end{equation}
Emphasize that the angles $\mathbf{\Delta}$ which specify the relative
orientation of the molecular frames introduced above are fixed and
known for any particular molecule. To illustrate the use of Eq.(\ref{MM}),
let as consider a perpendicular transition of a symmetric top molecule.
Then $\widetilde{G}_{kn}^{j}\equiv\delta_{kn}^{j}\widetilde{G}_{nn}^{j}$
and \begin{equation}
G_{00}^{j}(t)=\sum_{n=-j}^{j}D_{0n}^{j}(0,-\pi/2,0)\widetilde{G}_{nn}^{j}(t)D_{n0}^{j}(0,\pi/2,0).\label{MMS}\end{equation}
For the small angle rotational diffusion, for example, \begin{equation}
\widetilde{G}_{nn}^{j}=\exp\{-j(j+1)D_{\Vert}t-n^{2}(D_{\bot}-D_{\Vert})t\},\label{dif}\end{equation}
 $D_{\Vert}$ and $D_{\bot}$ being the corresponding diffusion coefficients.
\cite{ste63} Asymmetric top OCFs $G_{00}^{j}(t)$ within the  small
angle rotational diffusion equation can be computed, e.g., by the method described
in. \cite{pec76} 

For the sake of the further comparison, we also present the standard
formulas for the intensity of the polarized emission in ensemble measurements.
\cite{gor68,feo} Incorporating the cross-sections (\ref{Sa}) and
(\ref{SiEns}) into Eq.(\ref{I}) and  applying Eqs. (\ref{B1(t)})-(\ref{OCF1}), one obtains the standard result that the
averaged emission intensity is uniquely determined by the second-rank OCF of
the dipole moments involved, \begin{equation}
\left\langle I_{i}\right\rangle \sim1+\frac{4}{5}P_{2}(\mathbf{e}_{a}\mathbf{e}_{i})\left\langle P_{2}(\mathbf{\mu}_{a}\mathbf{\mu}(t)\right\rangle .\label{Ifl}\end{equation}
Here $P_{j}(x)$ are the $j$-rank Legendre polynomial and \begin{equation}
\left\langle P_{j}(\mathbf{\mu}_{a}\mathbf{\mu}(t)\right\rangle \equiv\sum_{m,m'=-j}^{j}D_{0m}^{j}(0,-\alpha_{\mu_{a}},-\beta_{\mu_{a}})G_{mm'}^{j}(t)D_{m0}^{j}(\alpha_{\mu},\beta_{\mu},0),\label{Leg}\end{equation}
 $ \alpha_{\mu_{a}},\beta_{\mu_{a}}$ and $\alpha_{\mu},\beta_{\mu}$ being the spherical angles which specify the orientation of the absorption ($\mathbf{\mu}_{a}$) and emission ($\mathbf{\mu}$) dipole moments in the molecular reference frame.

\section{Rank dependence of the single molecule signal}

Once the emission dipole moment $\mathbf{\mu}$ is pointed along the
$z$-axis of the molecular reference frame, then Eqs. (\ref{C(t)sim}) and (\ref{BD1})
determine the SM CF. The coefficients $A_{00}^{j}$ depend upon the
OCF rank $j$, the $NA$ angle $\chi$, and the relative orientation
$\varphi$ of the polarizers $\mathbf{e}_{1}$ and $\mathbf{e}_{2}$
($\varphi\equiv\arccos(\mathbf{e}_{1}\mathbf{e}_{2})$). Hereafter,
we denote these coefficients as $A^{j}(\chi,\varphi)$. The standard
choice of the orthogonal detection scheme corresponds to $A^{j}(\chi,\pi/2)$.
The term $A^{0}(\chi,\varphi)\equiv0$, since the isotropic component does not contribute
to dichroism (\ref{Xgen}). Furthermore, the symmetry of the D-functions
dictates that only the coefficients with even $j$ contribute into
Eq.(\ref{C(t)sim}). It can thus be recast into the form\begin{equation}
C(t)=\sum_{\sigma=1}^{\infty}A^{2\sigma}(\chi,\varphi)G^{2\sigma}(t).\label{C(t)ev}\end{equation}
 (we have dropped the $00$-subscripts for the clarity of notation,
and the CF is assumed to be normalized to unity, that is $C(0)=1$).
Formally speaking, the SM CF is expressed by the linear combination
of all even-rank OCFs.\cite{foot4} Their relative contributions are determined
by the weighting coefficients $A^{2\sigma}(\chi,\varphi)$, which
decrease rapidly with the OCF rank $j$, so that a few of them contribute
significantly into Eq.(\ref{C(t)ev}). 

Let us consider a special case of the orthogonal registration scheme
($\varphi=\pi/2$) first. If the $NA$ is negligibly small ($\chi=0$),
one gets the set of coefficients $A^{2\sigma}(0,\pi/2)$, which have been
calculated and analysed in. \cite{bas04,bou05} In this case, $A^{2}(0,\pi/2)=0.835$,
$A^{4}(0,\pi/2)=0.100$, etc., so that the contribution due to the second-rank
OCF yields more than $83$\%. In the opposite case of a high $NA$
($\chi=\pi/2$), as is clear from Eq.(\ref{Xsimp}), the second-rank
OCF contributes exclusively into CF (\ref{C(t)ev}), and $C(t)\sim\left\langle P_{2}(\mathbf{\mu\mu}(t)\right\rangle $.
Thus, apart from the numerical factor, the SM CF coincides with the
ensemble averaged anisotropy. Since the parameters $p$ and $q$ in
Eq.(\ref{SiSi}) increase monotonously with $\chi$ (see Fig. 2),
the significance of all coefficients $A^{2\sigma}(\chi,\pi/2)$ with $\sigma>1$
decreases with $\chi$. As has been shown in,\cite{bou05} for example, the
CF $C(t)$ corresponding to $NA=0.6$, $n=1.4$ coincides, practically, with $G^{2}(t)$. 

The analysis which has been carried out in Ref.\cite{bou05} and the above
considerations mean that, under some typical experimental conditions 
($\varphi=\pi/2$ and $\chi$ being close to $\pi/2$),
the SM CF $C(t)$ is indistinguishable from the second-rank OCF $G^{2}(t)$.
If one considers a complicated dependence of $C(t)$ on OCFs of various
ranks as a nuisance, which obscures interpretation of the measured
signal, then the use of the orthogonal detection scheme in conjunction
with high-$NA$ optics allows one to {}``measure'' the standard
second-rank OCF. On the other hand, Eq.(\ref{C(t)ev}) hints at a unique
opportunity to learn about higher-rank OCFs experimentally, even with
the use of high-$NA$ optics. We suggest that the non-orthogonal registration
scheme, when $\mathbf{e}_{1}$ and $\mathbf{e}_{2}$ are not perpendicular
to each other, makes this possible. 

To clarify the situation, let us consider the signal intensity (\ref{Ifl})
which is measured within the ensemble-averaged spectroscopy. The intensity
consists of the sum of the isotropic component, which contains
no information about molecular rotation, and the anisotropic component,
$\left\langle P_{2}(\mathbf{\mu}_{a}\mathbf{\mu}(t)\right\rangle $.
The effect of the detection scheme  is exclusively defined by the
numerical factor $P_{2}(\mathbf{e}_{a}\mathbf{e}_{i})$, which determines
the relative weights of the two components. Therefore, in order to
extract OCF $\left\langle P_{2}(\mathbf{\mu}_{a}\mathbf{\mu}(t)\right\rangle $
out of the signal, it is sufficient to perform measurements of the
emission intensities at any two different polarizations $\mathbf{e}_{1}$
and $\mathbf{e}_{2}$. A common practice is to use the magic angle
conditions ($P_{2}(\mathbf{e}_{a}\mathbf{e}_{i})=0$, $\mathbf{e}_{a}\mathbf{e}_{i}=1/\sqrt{3}$),
as well as parallel ($P_{2}(\mathbf{e}_{a}\mathbf{e}_{i})=1$, $\mathbf{e}_{a}\mathbf{e}_{i}=1$)
and perpendicular ($P_{2}(\mathbf{e}_{a}\mathbf{e}_{i})=-1/2$, $\mathbf{e}_{a}\mathbf{e}_{i}=0$)
detection schemes. 

The situation with the SM polarization spectroscopy is very different.
The SM CF (\ref{C(t)ev}) has weighted contributions from virtually
all even-ranked OCFs, and the weights themselves, $A^{2\sigma}(\chi,\varphi)$,
are clearly detection scheme dependent. The coefficients $A^{2\sigma}(\chi,\varphi)$
for several few first $\sigma$ are plotted in Fig. 3 for $\chi=0$
(small $NA$) and for $\chi=\pi/2$ (high $NA$). The parameters
$p$ and $q$ in Eq.(\ref{Xgen}) increase monotonously with $\chi$
(see Fig. 2). One thus expects that the coefficients $A^{2\sigma}(\chi,\varphi)$
transform gradually from those depicted in Fig. 3 in the upper panel to those depicted
in the lower panel, following the increase of $\chi$. 

Evidently, $A^{2}(\chi,\varphi)$-components dominate the signal
for any polarization scheme, so that the contribution due to the second
rank OCFs is the most significant. Both $A^{2}(0,\varphi)$ and $A^{2}(\pi/2,\varphi)$
reach their maxima at $\varphi=\pi/2$: $A^{2}(0,\varphi=\pi/2)=0.835$
and $A^{2}(\pi/2,\varphi=\pi/2)=1$. This means that the standard
orthogonal scheme reflects predominantly the decay of the second rank
OCF, since $84\%$ of the low-$NA$ signal and $100$\% of the high-$NA$
signal is determined by $G^{2}(t)$. If 
the angle $\varphi$ between the polarizations $\mathbf{e}_{1}$ and $\mathbf{e}_{2}$
decreases, then the contribution due to $A^{2}(0,\varphi)$ also 
dominates, but the higher rank OCFs start to contribute more and
more significantly. For a low-$NA$ signal, for example, the forth-rank
contribution achieves its maximum of $23$\% at $\varphi=\pi/8$ while
the second order contribution remains as high as $46$\%. The shapes
of the curves $A^{2\sigma}(0,\varphi)$ and $A^{2\sigma}(\pi/2,\varphi)$
are evidently different. Every coefficient $A^{2\sigma}(0,\varphi)$
with $\sigma>1$, as a function of $\varphi$, exhibits an (asymmetric)
bell-like shape  with a single maximum. $A^{2}(\pi/2,\varphi)$,
as well as $A^{2}(0,\varphi)$, increase monotonously with $\varphi$, 
while all $A^{2\sigma}(\pi/2,\varphi)$
with $\sigma>1$ decrease rapidly. An overall tendency can be described
as follows: the closer are $\mathbf{e}_{1}$ and $\mathbf{e}_{2}$
to each other, the more $A^{2\sigma}(\chi,\varphi)$ coefficients contribute
to the signal. This is clarified by Figs. 4, in which we present
the completeness parameters \begin{equation}
B^{2\xi}(\chi,\varphi)=\sum_{\sigma=1}^{\xi}A^{2\sigma}(\chi,\varphi)\label{Bksi}\end{equation}
 for few first $\xi$. As to the low-$NA$ detection, the first few
$\sigma$ are necessary to faithfully reproduce the signal for $\varphi$
close to $\pi/2$, while much more terms are necessary
for small $\varphi$. The same is also true for the high-$NA$ detection,
but the convergence is much more rapid. In general, the high-$NA$
detection is more second-rank OCF dominated than the low-$NA$ one.
We emphasize however that the high-$NA$ CF $C(t)$ coincides with
$G^{2}(t)$ in the case of the orthogonal detection scheme only ($\varphi=\pi/2$).
Otherwise, it also depends on OCFs of different even ranks. 

We conclude the present section with the following comment. On writing
the starting Eq.(\ref{X}) for the SM dichroism, we have tacitly
assumed that the emission intensities $I_{1}$ and $I_{2}$ scale
identically. To take into account a possible imbalance $\eta$ of
the channels, we can redefine the SM dichroism as $(I_{1}-\eta I_{2})/(I_{1}+\eta I_{2})$.
\cite{hoc04} We can then repeat the above analysis, taking into
account that the coefficients $A^{2\sigma}(\chi,\varphi)$ acquire
an additional $\eta$-dependence. In that case, for example, the isotropic
contribution is nonzero, $A^{0}(\chi,\varphi)\sim1-\eta\neq0$. If,
furthermore, there are several emission dipole moments,\cite{hoc03}
the above approach can also be generalized straightforwardly.

\section{Extraction of the high-rank orientational correlation functions from the single molecule signal}

As has been demonstrated in Sec.5, the non-orthogonal detection schemes
contain, potentially, more information on the high-rank OCFs than
the standard orthogonal schemes. The question thus arises if it is
possible to extract $G^{2\sigma}(t)$ with $\sigma>1$ from the SM
CF $C(t)$. To clarify the situation, we compute the signal within
the simplest model of molecular reorientation, the spherical top small
angle diffusion. Within this model, \cite{ste63} the OCFs are described
by the exponential formula 
\begin{equation}
G_{00}^{j}(t)=\widetilde{G}_{00}^{j}(t)=\exp\{-j(j+1)D_{\Vert}t\}.\label{dif2}
\end{equation}
We have chosen this model since it is most frequently applied for
the interpretation of SM CFs. 

Let us consider the small-$NA$ detection first (Fig. 5). $C(t)$
calculated within the orthogonal detection scheme ($\varphi=\pi/2$)
and within the scheme with $\varphi=\pi/8$ (this angle provides the
maximum for the $G^{4}(t)$-contribution, see Fig. 3, upper panel) are seen to be
markedly different, both mutually and from $G^{2}(t)$. On the other
hand, the SM CF with $\varphi=\pi/2$ is rather close to $G^{2}(t)$, and
the slopes of both SM CFs, as well as the slope of $G^{2}(t)$, are almost
the same. This is totally understandable, since (i) $84$\% of the
$\varphi=\pi/2$ CF and $46\%$ of its $\varphi=\pi/8$ counterpart
are determined by $G^{2}(t)$ and (ii) the higher-rank OCFs decay
much more rapidly than those with $j=2$. 

Theoretically speaking, the procedure of the extraction of the high-rank
OCFs from $C(t)$ is straightforward. One can perform several (say,
$N$) measurements at $N$ different detection angles $\varphi$, truncate
the number of summations in Eq.(\ref{C(t)ev}) by $N$, and solve
the corresponding system of linear equations for $G^{2\sigma}(t)$,
$\sigma=1\div N$. Such a procedure, however, can hardly be feasible
in practice. There exists a cruder, but much more robust procedure
which is exemplified by Fig. 5. We can calculate the quantity \begin{equation}
\overline{C}(t)=C(t,\varphi=\pi/8)-\frac{A^{2}(0,\pi/8)}{A^{2}(0,\pi/2)}C(t,\varphi=\pi/2).\label{Cbar}\end{equation}
 Evidently, $\overline{C}(t)$ represents a weighted sum of the OCFs
of the rank $j=4$ and higher. Since the higher-rank contributions
into $\overline{C}(t)$ decrease rapidly with $\sigma$ (Fig. 3)
one expects that $\overline{C}(t)$ is determined, predominantly,
by $G^{4}(t)$. This qualitative expectation is corroborated by Fig.
5. Evidently, $\overline{C}(t)$ and $G^{4}(t)$ do not coincide but,
as has been explained above, their slopes are virtually the same.

The procedure described above can also be applied to the high-$NA$
detection, see Fig. 6. Furthermore, the situation is much more fortunate
in this case, since the orthogonal CF is now solely determined by
the second rank OCF $G^{2}(t)$ ($A^{2}(\pi/2,\pi/2)=1$, see also
Fig. 3). It is interesting that CFs $C(t,\varphi=\pi/4)$ and $C(t,\varphi=\pi/2)$,
which are presented in the Fig. 6, are rather close to each other,
since both of them are predominantly determined by $G^{2}(t)$. However,
the CF\begin{equation}
\overline{C}(t)=C(t,\varphi=\pi/4)-A^{2}(\pi/2,\pi/4)C(t,\varphi=\pi/2)\label{Cbar1}\end{equation}
is almost indistinguishable from $G^{4}(t)$. 

The above results demonstrate that the proposed (or similar) procedure of the extraction of high-order OCFs from the SM CF $C(t)$ is useful
and robust, both for high- and low $NA$ objectives.  Once the higher order OCFs are available, one can get  valuable information on the dynamics of the SM reorientation. For example, let us suppose that orientational relaxation proceeds exponentially. Then, if the second-rank OCF is available only, one can extract an effective time of the OCF decay, but cannot learn anything about rotational dynamics. If both $G^{2}(t)$ and $G^{4}(t)$ are known, one then can calculate the ratio of their relaxation times. This quantity, being very model-specific, allows one to discriminate between different reorientation mechanisms. For the small-angle diffusion model, the ratio equals to $10/3$ (see Eq.(\ref{dif2})).  For the jump diffusion model, it is close to $1$.\cite{val73,cuk72,cuk74} More sophisticated approaches to the orientational relaxation \cite{McCl77,BurTe,kos06,wan80,sza84,key72,ste84a,kiv88} also predict that the comparison of the behavior of OCFs of different ranks makes it possible to identify the underling mechanisms of orientational relaxation. 


Before concluding the present section, we wish to discuss possible causes of the deviation of $C(t)$
from the exponential form (see also Refs. \cite{bas04,bou05}). There can
be two fundamentally different groups of reasons. First, $C(t)$ is
described by the weighted sum of the even-rank OCFs, Eq.(\ref{C(t)ev}).
Despite the second-rank OCF contributes predominately into $C(t)$,
the contributions due to the higher-rank OCFs cannot be neglected,
in general. It is only for a high-$NA$ objective and close-to-orthogonal
detection scheme that these higher order contributions are vanishingly
small and $C(t)\sim G^{2}(t)$. Second, the OCFs (including the second-rank OCF)
can be non-exponential due to a variety of reasons. (i). Molecular
rotation is not necessarily diffusive. While the jump diffusion model
predicts exponentially decaying OCFs \cite{val73,cuk72,cuk74} (although
their rank-dependence differs from that given by Eq.(\ref{dif})), the restricted
diffusion model, \cite{wan80,sza84} the diffusion-equation-with-memory
models and other memory function approaches \cite{key72,ste84a,kiv88}
predict the OCFs of a spherical molecule to be described by the sum
of several (in general, complex) exponentials. More sophisticated approaches
deliver, of course, more complex OCFs. On the other hand, the so-called
inertial effects, which induce highly non-exponential behavior of OCFs,
\cite{McCl77,BurTe,kos06} are irrelevant for the SM spectroscopy since
they manifest themselves on a time scale which is much faster than
the time resolution of typical SM experiments. (ii). A deviation of
the molecular shape from spherical complicates molecular reorientation
even in the small angle diffusion limit. The second-rank OCF $G_{00}^{2}(t)$
of an asymmetric top, for example, is described by the sum of two
exponentials. \cite{ste63,fav60} (iii). Of relevance is the direction
of the emission dipole moment $\mathbf{\mu}$ in the molecular reference frame
(see Eqs. (\ref{MM}) and (\ref{MMS})). For example, if $\mathbf{\mu}$
is parallel to the axis of the linear or symmetric rotor, the corresponding
second-rank OCF is mono-exponential. If $\mathbf{\mu}$ is perpendicular
to the molecular axis, then the corresponding OCF is two-exponential (see Eq.(\ref{dif})).
(iv). Finally, internal rotations can also cause deviations from exponentiality.
\cite{sza84}

\section{Conclusion}

We have developed an approach for the calculation of the SM
CF $C(t)$, Eq.(\ref{C(t)}), in the general case of asymmetric
top molecules. The key dynamic quantities of our analysis are the
even-rank OCFs $G^{j}(t)$ (\ref{OCF1}), the weighted sum of which
constitutes $C(t)$. The OCFs can either be evaluated within any model
of molecular reorientation available in the literature \cite{McCl77,BurTe,kos06}
or simulated on a computer. \cite{AlTi} We have demonstrated that
the use of non-orthogonal schemes for the detection of SM polarization
responses allows one to manipulate the weighting coefficients in the
expansion of $C(t)$ on OCFs. Thus a valuable information about the
OCFs of the rank higher than second can be extracted from the SM CF
$C(t)$. Until recently, such an information was available only through
computer simulations and/or model calculations. Neither the corresponding
information is accessible within the ensemble-averaged spectroscopy,
in which one ``measures'' exclusively the second-rank OCF (\ref{Ifl}).

\clearpage

\begin{center}
{\bf FIGURE CAPTIONS: }
\end{center}

{\sf  \bf FIGURE 1}:
Schematic representation of the SM polarization experiment. 

{\sf  \bf FIGURE 2}:
The $NA$-angle dependence of the coefficients $p$
(dashed line) and $q$ (full line). 

{\sf  \bf FIGURE 3}:
The expansion coefficients $A^{2\sigma}(0,\varphi)$  (low-$NA$ objective, upper panel) and  $A^{2\sigma}(\pi/2,\varphi)$ (high-$NA$ objective, lower panel) 
as functions of the angle $\varphi$ between the polarizers.
From top to bottom, the curves correspond to $\sigma=1\div8$ (upper panel) and 
$\sigma=1\div6$ (lower panel).

{\sf  \bf FIGURE 4}:
The completeness parameters $B^{2\sigma}(0,\varphi)$  (low-$NA$ objective, upper panel) and  $B^{2\sigma}(\pi/2,\varphi)$ (high-$NA$ objective, lower panel)
as functions
of the angle $\varphi$ between the polarizers.
From top to bottom, the curves correspond to $\sigma=1\div8$ (upper panel) and 
$\sigma=1\div4$ (lower panel).

{\sf  \bf FIGURE 5}:
Extraction of the higher-order OCFs from the single molecule
CF $C(t)$ (\ref{C(t)ev}) in the case of a low-$NA$ objective ($\chi=0$).
The upper solid curve depicts the the second-rank OCF $G^{2}(t)$
and the lower solid curve depicts the forth-rank OCF $G^{4}(t)$.
The upper dashed curve shows the single molecule CF $C(t)$ calculated
for $\varphi=\pi/8$ and the lower dashed curve shows the single molecule
CF $C(t)$ calculated for $\varphi=\pi/2$. The dotted curve depicts
the forth-rank OCF, which is approximately extracted from the two single molecule CFs (see text for details).
All the CFs are calculated within the small angle rotational diffusion
model for a spherical top, $D_{\Vert}=0.1$; $t$ and $D_{\Vert}^{-1}$
are given in arbitrary dimensionless units. 

{\sf \bf FIGURE 6}:
Extraction of the higher-order OCFs from the single molecule
CF $C(t)$ (\ref{C(t)ev}) in the case of a high-$NA$ objective ($\chi=\pi/2$).
The upper curves depict the second-rank OCF $G^{2}(t)$ (solid line)
and the single molecule CF $C(t)$ calculated for $\varphi=\pi/4$
(dashed line). The lower curves show the exact forth-rank OCF $G^{4}(t)$
(solid line) and its counterpart extracted from CF $C(t)$ (dashed
line, see text for details). All the CFs are calculated within the
small angle rotational diffusion model for a spherical top, $D_{\Vert}=0.1$;
$t$ and $D_{\Vert}^{-1}$ are given in arbitrary dimensionless units. 

\clearpage

\begin{figure}
\includegraphics[keepaspectratio,totalheight=10cm]{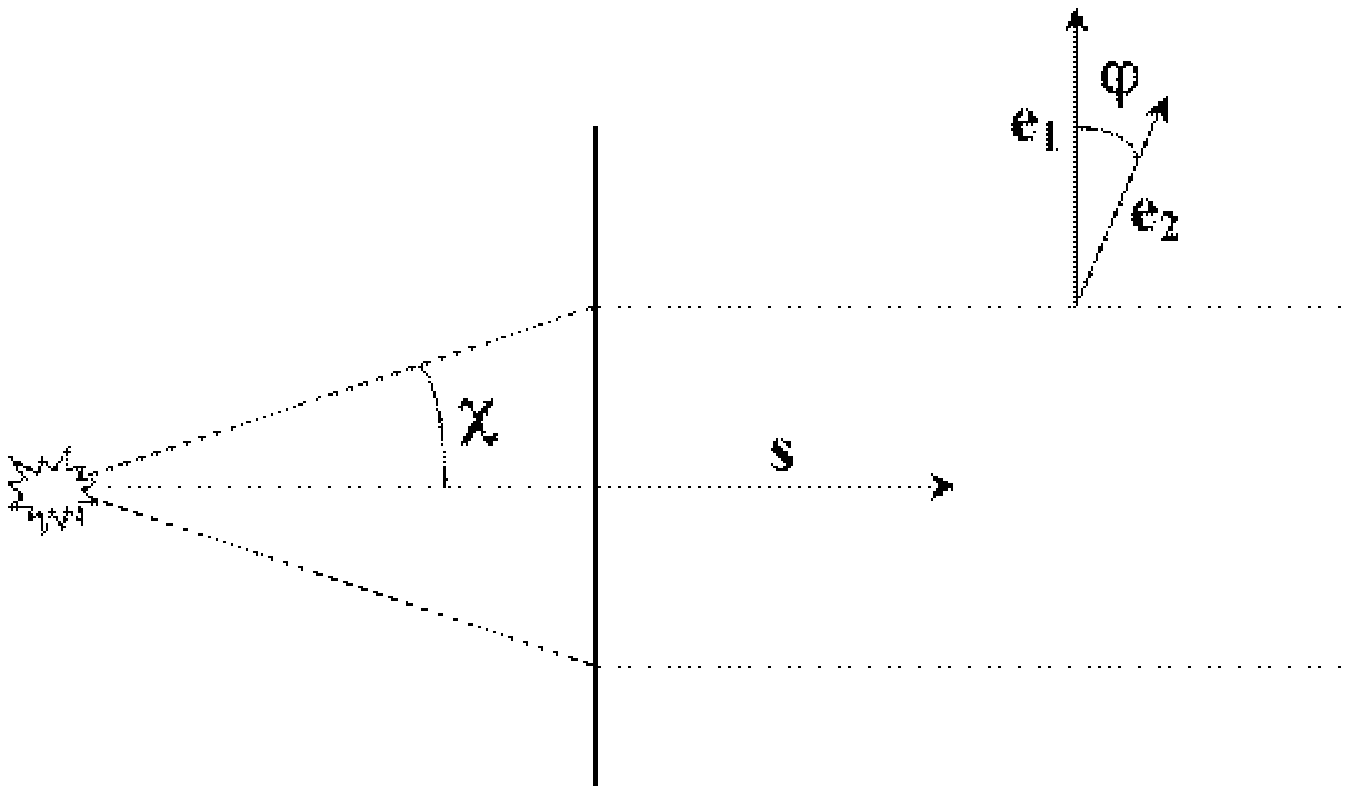}
\caption{
}
\end{figure}

\clearpage
\begin{figure}
\includegraphics[keepaspectratio,totalheight=10cm]{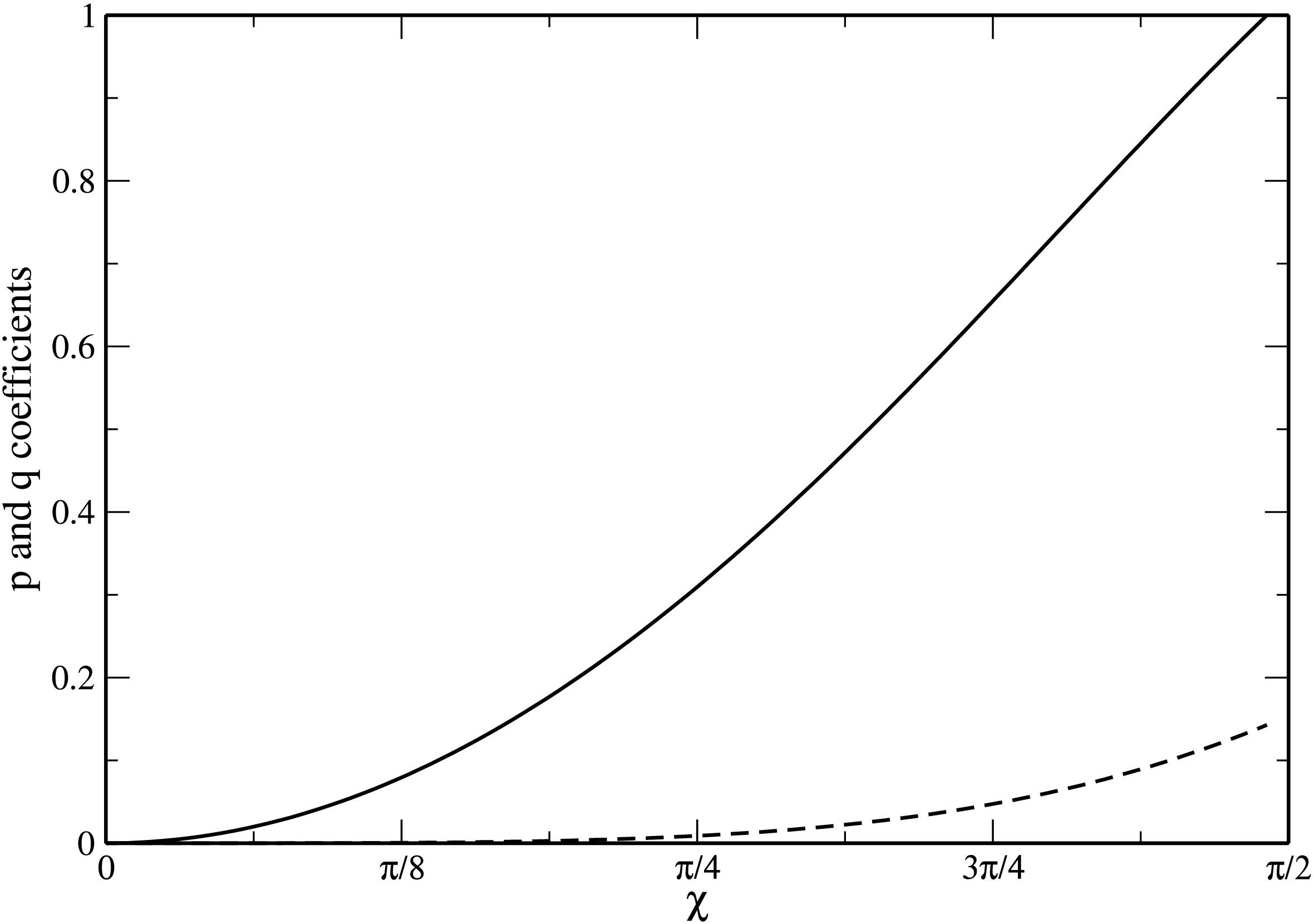}
\caption{ 
}
\end{figure}

\clearpage

\begin{figure}
\includegraphics[keepaspectratio,totalheight=10cm]{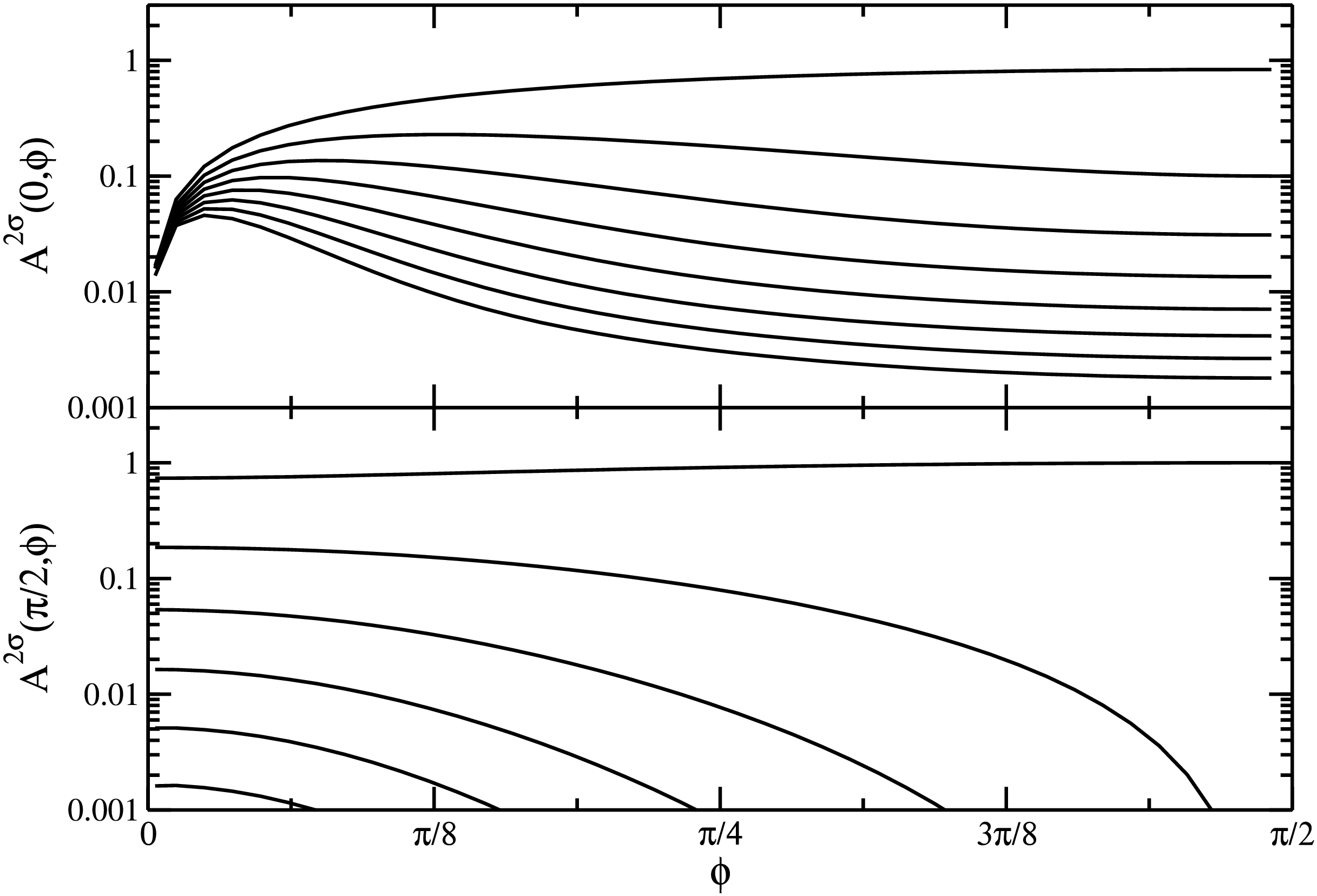}
\caption{ 
}
\end{figure}

\clearpage

\begin{figure}
\includegraphics[keepaspectratio,totalheight=10cm]{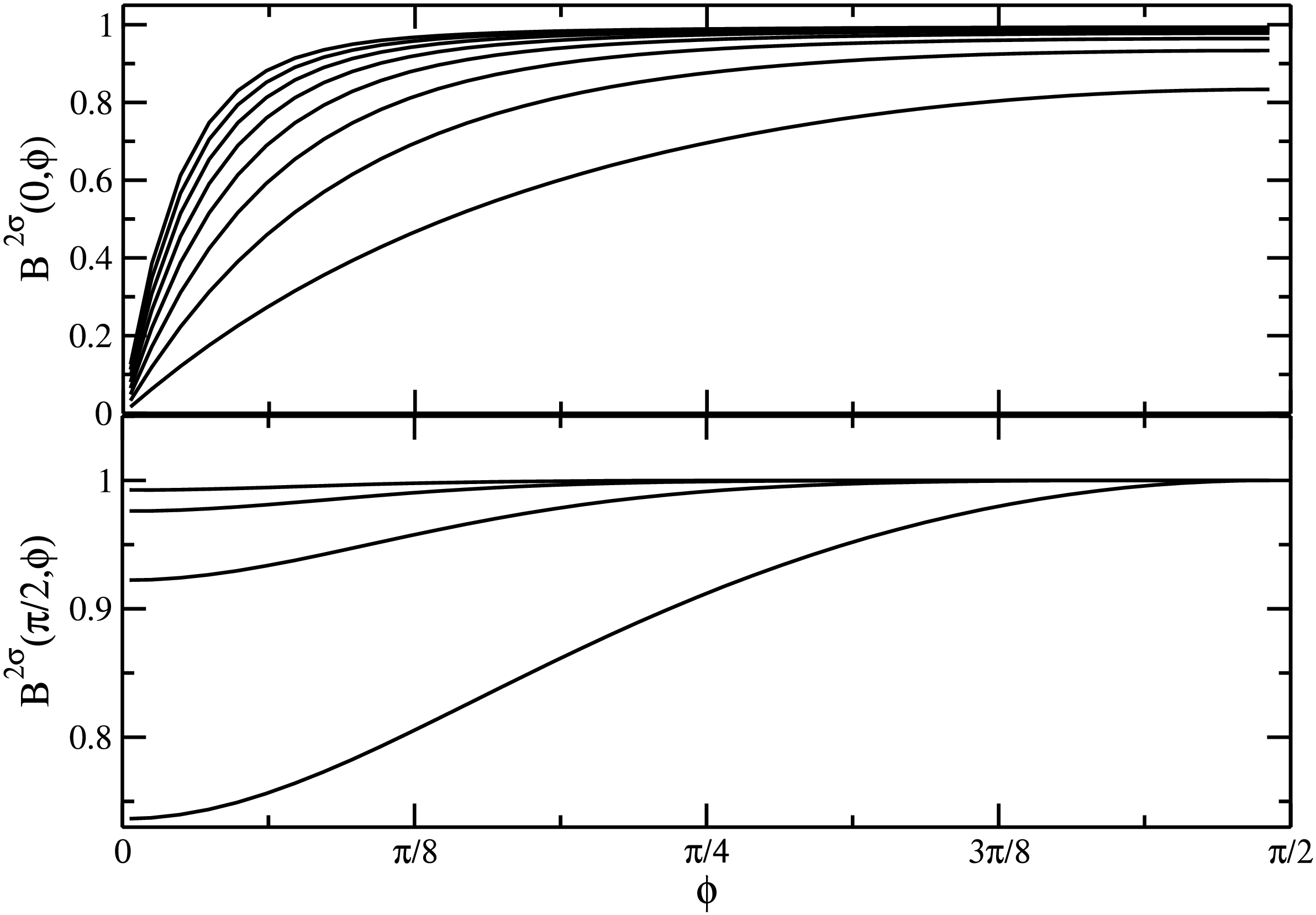}
\caption{ 
}
\end{figure}

\clearpage

\begin{figure}
\includegraphics[keepaspectratio,totalheight=10cm]{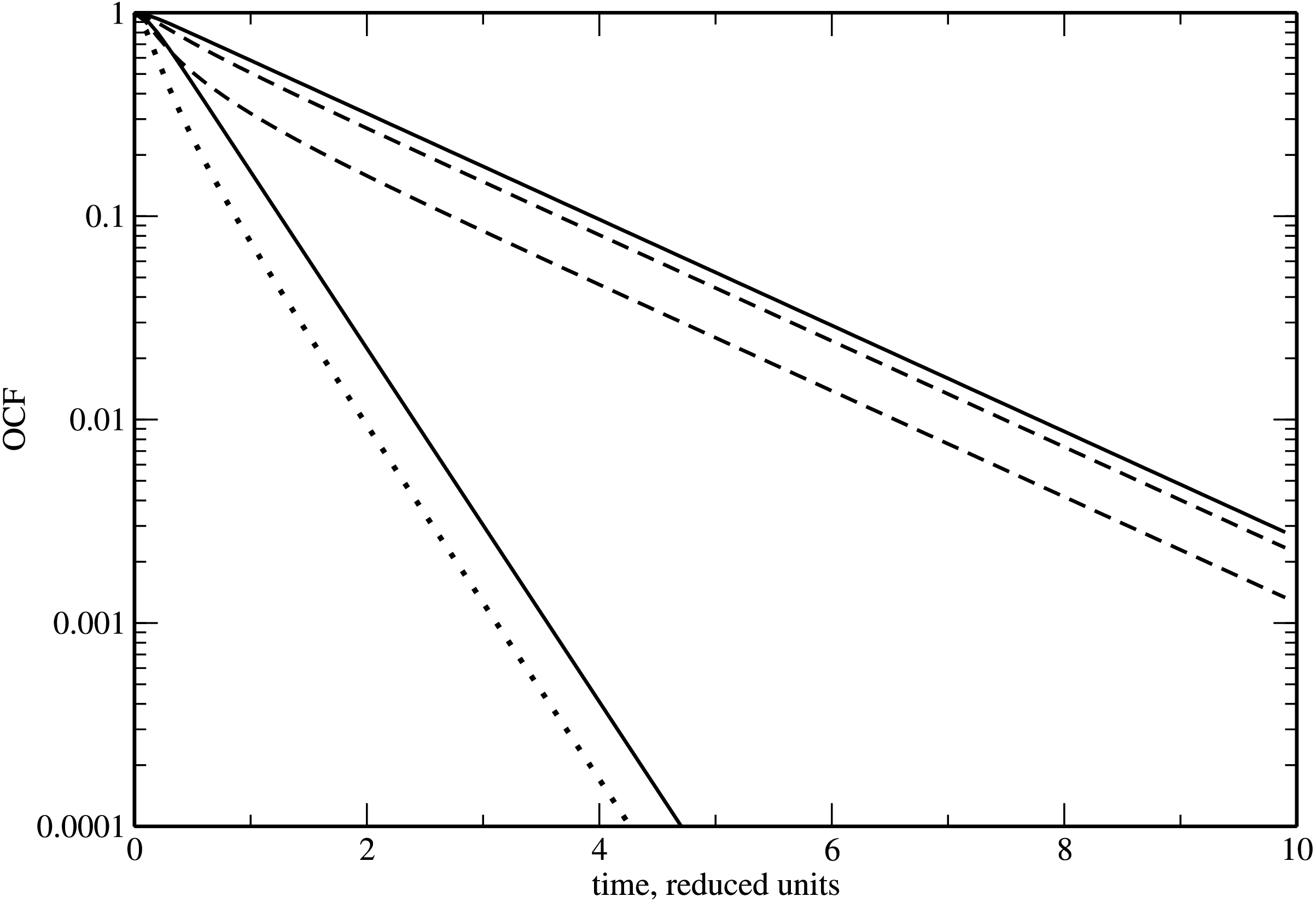}
\caption{ 
}
\end{figure}

\clearpage

\begin{figure}
\includegraphics[keepaspectratio,totalheight=10cm]{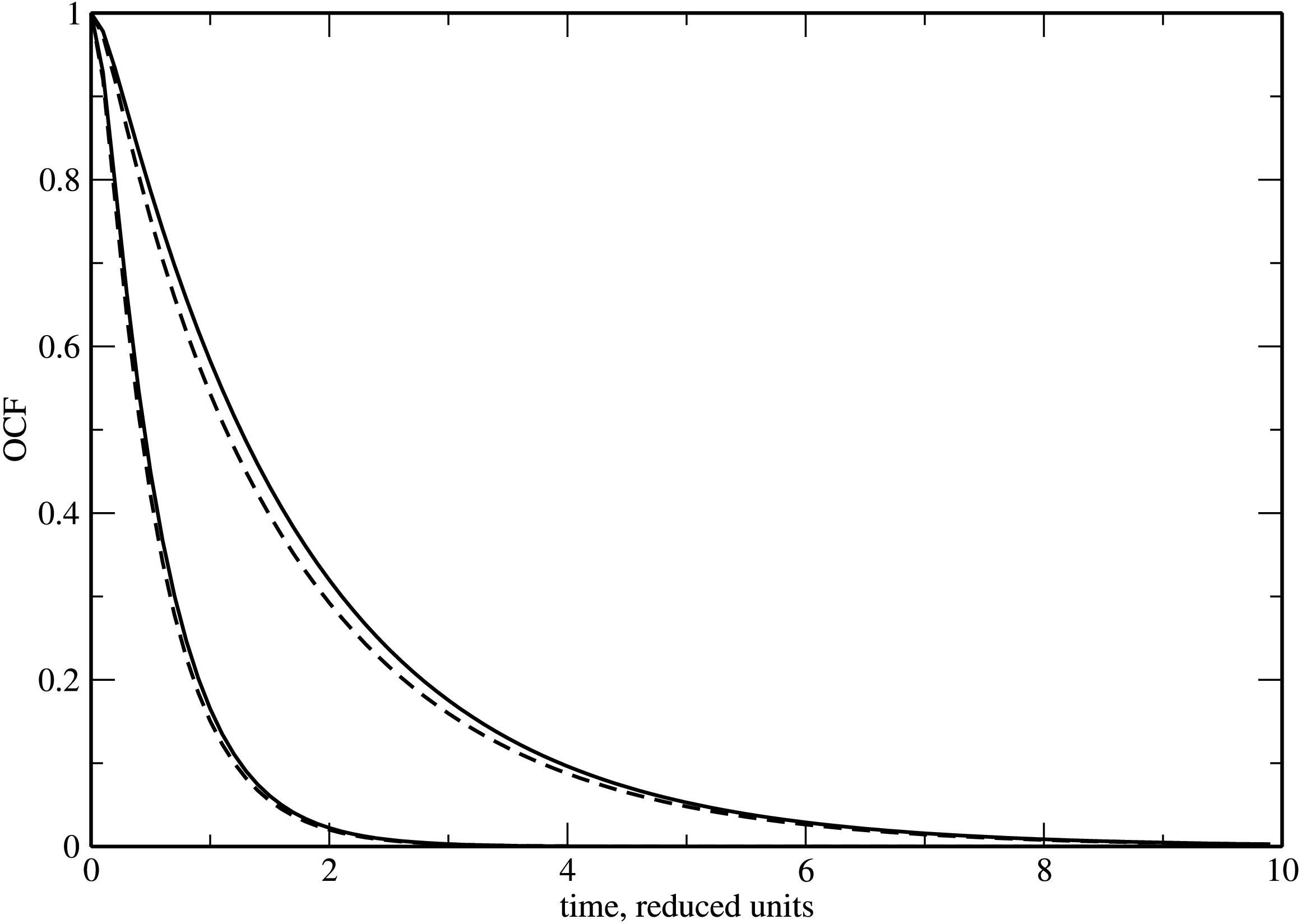}
\caption{ 
}
\end{figure}


\begin{thebibliography}{20}
\bibitem{foot3}If, prior to the collecting objective, the emitted photons pass
through an interface, then Eq.(\ref{eps}) can be generalized as
described, e.g., in Ref. \cite{end03}
\bibitem{foot1}Other SM-relevant polarization CFs (see, e.g., Ref. \cite{ha04}) can be
calculated very similarly.
\end{thebibliography}
\end{document}